\documentclass[12pt,preprint]{aastex}
\shorttitle{Selection Effects on the $\theta_j-z$ dependence of GRBs}\shortauthors{Lu et al.}
\usepackage{amsmath}
\slugcomment{}

\begin{document}

\title{Selection Effects on the Observed Redshift Dependence of GRB Jet Opening Angles}
\author{Rui-Jing Lu\altaffilmark{1}, Jun-Jie Wei\altaffilmark{1}, Shu-Fu Qin\altaffilmark{2} and En-Wei Liang\altaffilmark{*1,2}}
\altaffiltext{1}{Department of Physics and GXU-NAOC Center for Astrophysics and
Space Sciences, Guangxi University, Nanning 530004, China; lew@gxu.edu.cn}
\altaffiltext{2}{The National Astronomical Observatories, Chinese Academy of
Sciences, Beijing 100012, China}

\begin{abstract}
Apparent redshift dependence of the jet opening angles ($\theta_{\rm
j}$) of gamma-ray bursts (GRBs) is observed from current GRB sample.
We investigate whether this dependence can be explained with
instrumental selection effects and observational biases by a
bootstrapping method. Assuming that (1) the GRB rate follows the
star formation history and the cosmic metallicity history and (2)
the intrinsic distributions of the jet-corrected luminosity ($L_{\rm
\gamma}$) and $\theta_{\rm j}$ are a Gaussian or a power-law
function, we generate a mock {\em Swift}/BAT sample by considering
various instrumental selection effects, including the flux threshold
and the trigger probability of BAT, the probabilities of a GRB jet
pointing to the instrument solid angle and the probability of
redshift measurement. Our results well reproduce the observed
$\theta_{\rm j}-z$ dependence. We find that in case of
$L_{\gamma}\propto \theta_{\rm j}^2$ good consistency between the
mock and observed samples can be obtained, indicating that both
$L_{\rm \gamma}$ and $\theta_{\rm j}$ are degenerate for a
flux-limited sample. The parameter set $(L_{\rm \gamma}, \theta_{\rm
j})=(4.9\times 10^{49} \rm {erg\ s}^{-1},\ 0.054 {rad} )$ gives the
best consistency for the current {\em Swift} GRB sample. Considering
the beaming effect, the derived intrinsic local GRB rate accordingly
is $2.85\times 10^2$ Gpc$^{-3}$ yr$^{-1}$, inferring that $\sim
0.59\%$ of Type Ib/c SNe may be accompanied by a GRB.
\end{abstract}

\section{Introduction\label{sec:intro}}
Gamma-ray bursts (GRBs) are the most luminous explosions in the
Universe. They are detected from deep to our local Universe (Zhang
\& Meszaros 2004). The measured redshifts of current GRB sample
range from 0.0085 (GRB 980425; Tinney et al. 1998) to 8.2 (GRB
090423; Tanvir et al. 2009)\footnote{A photometric redshift of 9.4
for GRB 090429B was reported by Cucchiara et al. (2011)}. It is
generally believed that the births of long GRBs follow the star
formation history of the universe (e.g., Totani 1997; Paczynski
1998; Bromm \& Loeb 2002; Lin et al. 2004), and they may be
promising probes for cosmology and galaxy evolution (e.g., Dai et
al. 2004; Ghirlanda et al. 2004; Liang et al. 2005; Y\"{u}ksel \&
Kistler 2007; Kistler et al. 2008, 2009; Salvaterra et al. 2008;
Wang \& Dai 2009). However, it is still unclear if GRBs experience
any sort of cosmic evolution.

With high sensitivity of the burst alert telescope (BAT) and the
promptly slewing capability of the narrow-field X-ray telescope
(XRT) onboard the {\em Swift} satellite(Gehrels et al. 2004), the
number of measured high-$z$ GRBs has increased rapidly over the last
6.5 years. It is found that long GRBs do not follow star formation
rate (SFR) unbiasedly (e.g., Kistler et al. 2008). The observed GRB
rate is higher than the SFR at $z\sim 4$, which may be due to that
the GRB rate is also related to the comic metallicity history
(Kistler et al. 2008; Li 2008; Qin et al. 2010; Virgili et al. 2011)
or due to a higher efficiency of long GRB production rate by massive
stars at high-$z$ (Daigne et al. 2006). It was also suggested that
the GRB luminosity may evolve with redshift, being more luminous in
the past, which is simplified as $L\propto L_0(1+z)^{\delta}$, where
$L_0$ is the luminosity of local GRBs (Lloyd-Ronning, Fryer \&
Ramirez-Ruiz 2002; Firmani et al. 2004; Kocevski \& Liang 2006;
Salvaterra et al. 2008, 2009).

It is believed that GRBs should be collimated into narrow jets
(Rhoads 1997). The collimation of GRB jet is critical to understand
the physics of GRBs, such as their central engine, intrinsic GRB
rate, and true gamma-ray energy explosion (e.g., Zhang et al. 2004).
The observed steepening with a decay slope of $\sim 2$ in the late
afterglow lightcurves is possibly due to the jet effect, in which a
break would happen when the bulk Lorentz factor of the outflow
($\Gamma$) slows down to $\Gamma<\theta^{-1}_{j}$, where
$\theta_{\rm j}$ is the half opening angle of the jet (Rhoads 1997;
Sari et al. 1999). With the observed jet break time $t_{\rm j}$, the
$\theta_{\rm j}$ can be calculated (Rhoads 1997; Chevalier \& Li
2000). Frail et al. (2001) derived the geometrically-corrected
gamma-ray energy ($E_\gamma$) of a small GRB sample and proposed an
idea of a standard energy reservoir in GRB jets based on the narrow
distribution of $E_\gamma$ (see also Berger et al. 2003 for the same
conclusion for the jet-corrected luminosity ($L_{\rm \gamma}$).
However, this result seems to be a sample selection effect. The
$E_{\rm \gamma}$ distribution is significantly broadened with the
current GRB sample, although it is still log-normal (Liang et al.
2008; Racusin et al. 2010). It was also proposed that GRB jet
opening angles may experience similar kind of evolution as GRB
luminosity (Wei et al. 2003). Estimated the $\theta_{\rm j}$ of the
BATSE GRB sample with an empirical relation based on the Yonetoku
and Ghirlanda relations (Ghirlanda et al. 2004; Yonetuku et al.
2004), Yonetoku et al. (2005) argued that $\theta_{\rm j}$ evolves
as $\theta_{j}\propto(1+z)^{-0.45}$.

The evolution features of GRB luminosity and jet opening angle as
well as the GRB jet energy distribution are based on the statistics
for the current observational samples. However, as well known,
observational samples highly suffers instrumental selection effects
and observations biases (Bloom 2003). It is difficult to reveal the
intrinsic distributions and evolution features of GRBs from these
samples. For example, a GRB with a peak photon flux lower than an
instrumental threshold cannot trigger the instrument. Even the flux
is over the threshold, the trigger probability of a GRB strongly
depends on its photon flux (Qin et al. 2010). The probability of a
GRB jet pointing to the solid angle of an instrument is different
for GRBs with different $\theta_{\rm j}$. In addition, $E_{\rm
\gamma}$ (or $L_{\rm \gamma}$) and $\theta_{\rm j}$ are not direct
observable. The derivations of these parameters are mainly based on
the observed flux, redshift, and $t_{\rm j}$. However, both $E_{\rm
\gamma}$ (or $L_{\rm \gamma}$) and $\theta_{\rm j}$ are degenerate
for a flux-limited sample. This means that an intrinsically dimmer
GRB (with lower $E_{\rm \gamma}$ or $L_{\rm \gamma}$) may be
detected by a given instrument in case of a smaller $\theta_{\rm
j}$.

In this paper we investigate whether the observed redshift
dependences of jet opening angles and isotropic gamma-ray luminosity
of long GRBs can be explained with instrumental selection effects
and observational biases. Based on our results we also estimate
the local long GRB rate by considering the beaming effect. We search
for GRBs whose jet break times in afterglow lightcurves are reported
in literature. The data are reported in Section2. Our models and
results for {\em Swift}/BAT sample are reported in
section 3. Conclusions and Discussion are presented in Section 4.
Throughout we use cosmological parameters $H_{0}=71 {\rm km\ s}^{-1}
\rm{Mpc}^{-1}$, $\Omega_{M}=0.3$, $\Omega_{\Lambda}=0.7$.

\section{Data\label{sec:data}}
\begin{figure}
\includegraphics[angle=0,scale=0.6]{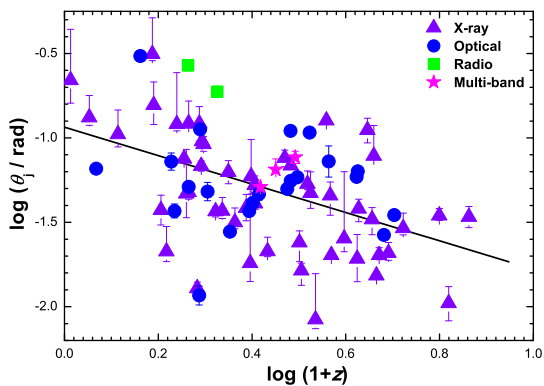}
\caption{Jet opening angles as a function of redshift for the 77
GRBs in Table 1.  The GRBs with $t_{\rm j}$ measuring at different
bands are marked with different symbols as indicating in the legend.
The solid line is the best fit to all data.}\label{thetaj}
\end{figure}

Our sample includes all GRBs whose jet break times ($t_{\rm j}$) are
measured in the radio, optical, and X-ray afterglow lightcurves,
regardless the $t_{\rm j}$ are achromatic, or just detected only in
one band. Note that a tentative achromatic jet breaks in the radio,
optical and X-ray afterglow lightcurves are available for three
GRBs. We take the $t_{\rm j}$ of these GRBs measured in the optical
afterglow lightcurves. Seventy-seven GRBs are included in our
sample. They are summarized in Table 1. Since different cosmological
parameters and medium density surrounding the bursts are used in
calculations of $\theta_{\rm j}$ in literature, we re-derive $E_{\rm
iso}$, $\theta_{\rm j}$, and $E_{\rm \gamma}$  in a consistent
method with the same cosmological parameters and the same jet model
for a constant medium density (Rhoads 1997, Sari et al. 1999, Frail
et al. 2001)
\begin{equation}\label{jet}
\begin{split}
\theta_{\rm j}=0.057 {\rm rad}(\frac{t_{j}}{1\ \rm
day})^{3/8}(\frac{1+z}{2})^{-3/8}(\frac{E_{\rm iso}}{10^{53}\ \rm
ergs})^{-1/8}(\frac{\eta_{\rm \gamma}}{0.2})^{1/8}(\frac{n}{0.1\ \rm
cm^{-3}})^{1/8},
\end{split}
\end{equation}
where  $n$ is the ambient medium density in unit of cm$^{-3}$ and
$\eta_{\rm \gamma}$ is the efficiency of the fireball in converting
the energy in the ejecta into gamma-rays. We take $\eta_{\rm
\gamma}=0.2$ and $n\sim 0.1$ cm$^{-3}$ in our calculations. The
$E_{\rm iso}$ is calculated with
\begin{equation}
E_{\rm iso}=\frac{4\pi D^{2}_{L}S_\gamma}{(1+z)}k,
\end{equation}
where $S_\gamma$ is the observed gamma-ray fluence, $D_L$ is the
luminosity distance at redshift $z$, and $k$ is a factor to correct
the observed gamma-ray energy in a given band pass to a broad band
(i.e., $1-10^4$ keV in this analysis) with the observed GRB spectra
(Bloom et al. 2001). We collect the spectral parameters for the
burst in our sample from literature.  It is well known that the GRB
spectrum is well fit with the Band function (Band et al. 1993).
Since the narrowness of the {\em Swift}/BAT band, the spectra of
most {\em Swift} GRBs in our sample are adequately fit with a single
power-law, $N\propto E^{-\Gamma}$ (Zhang et al. 2007; Sakamoto et al. 2009).  If the
parameters of the Band function are unavailable for the {\em Swift}
GRBs, we use an empirical relation between the peak energy of the
$\nu f_\nu$ spectrum and the photon index $\Gamma$, i.e. $\log
E_{p}=(2.76\pm0.07)-(3.61\pm0.26)\log \Gamma$ (Zhang et al. 2007), to
estimate the  $E_{p}$  and take the low and high photon indices as
$\alpha=-1.1$ and $\beta=-2.2$ (Preece et al. 2000; Kaneko et
al. 2006). The geometrically-corrected gamma-ray energy of a GRB jet
can be obtained with
\begin{equation}\label{Eiso}
E_{\gamma}=(1-\cos\theta_{\rm j})E_{\rm iso}.
\end{equation}
The derived $\theta_{\rm j}$ and $E_{\gamma}$ are reported in Table
1. It is found that their distributions are log-normal, as shown in
Figure \ref{DJEr}. The fits with a Gaussian function yield
$\log(\theta_{\rm j}/\rm rad)=(-1.31\pm 0.24) $ ($1\sigma$) and
$\log (E_{\gamma}/{\rm erg})=(50.07\pm 0.91)$ ($1\sigma$),
respectively. The $\theta_{\rm j}$ as a function of $z$ is shown in
Figure \ref{thetaj}. A tentative anti-correlation is observed. The
best fit gives
\begin{equation}\label{SBPL}
\log \theta_{\rm j}=(-0.90 \pm 0.09)-(0.94 \pm 0.19)\log (1+z),
\end{equation}
with a Spearman correlation coefficient of 0.55, a standard deviation of
0.30, and a chance probability $p<10^{-4}$ for $N=77$.

\section{Analysis of the Instrumental Selection Effects}
It is unclear whether the observed $\theta_{\rm j}-z$ dependence
shown above is due to observational selection effects or due to an
intrinsic cosmological evolution feature of GRBs. In this section,
we make bootstrap analysis of the instrumental selection effects on
this dependence.
\subsection{Models}
The number density of GRBs at redshift $z\sim z+dz$ is given by
\begin{equation}\label{density}
n(z,L_{\rm \gamma})=\frac{dN}{dzdL_{\rm \gamma}
dT}=\frac{R_{GRB}(z)}{1+z}\frac{dV(z)}{dz}\Phi(L_{\rm \gamma}),
\end{equation}
where $R_{\rm GRB}(z)$ is the co-moving GRB rate as a function of
$z$, the factor $(1+z)^{-1}$ accounts for the cosmological time
dilation of the observed rate, $dT$ is the time interval,
$\Phi(L_{\rm \gamma})$ is the luminosity function of GRB jets and
$dV(z)/dz$ is the co-moving volume element. We assume $R_{\rm GRB}$
traces the star formation rate and metallicity history, reading as
(Kistler et al. 2008; Li 2008; Qin et al. 2010; Virgili et al. 2011)
\begin{equation}\label{rate}
R_{\rm GRB}(z)=\rho_0 R_{\rm SFR}(z)\Theta(\epsilon,z),
\end{equation}
where $\rho_0$ is the local GRB rate, $\Theta(\epsilon,z)$ is the
fractional mass density belonging to metallicity below $\epsilon
Z_{\odot}$ at a given $z$ ($Z_{\odot}$ is the solar metal abundance)
and $\epsilon$ is determined by the metallicity threshold for the
production of GRBs. In our analysis, the $R_{\rm SFR}(z)$ is taken
as
\begin{equation} \label{SFR}
R_{\rm SFR}(z)\propto
\begin{cases}
(1+z)^{3.44}, \quad\quad \quad  z \leq z_{ \rm peak} \\
(1+z_{\rm peak})^{3.44},\quad \quad  z > z_{\rm peak}
\end{cases},
\end{equation}
where $z_{\rm peak}=1$ (Rowan-Robinson 1999). The
$\Theta(\epsilon,z)$ is parameterized as (Hopkins \& Beacom 2006;
Langer \& Norman 2006)
\begin{equation} \label{meta}
\Theta (\epsilon, z)= \frac{\hat{\Gamma}(\alpha+2,\epsilon^\beta
10^{0.15\beta z} )}{\Gamma(\alpha+2)},
\end{equation}
where $\alpha$ and $\beta$ are the low and high spectral indices of
the prompt gamma-rays, $\hat{\Gamma}(a,x)$ and $\Gamma(x)$ are the
incomplete and complete gamma function (Langer \& Norman 2006;
Kistler et al. 2008). We adopt the jet-corrected luminosity function
$\Phi(L_{\rm \gamma})$ as a log-normal distribution function
(Matsubayashi et al. 2005),
\begin{equation}\label{DLj}
\Phi(L_{\rm \gamma})=\frac{A_{\rm L}}{\sqrt{2\pi}\sigma_{L_{\rm
\gamma}}}\exp[-\frac{(\log L_{\rm \gamma}-\log L_{\rm
c})^{2}}{2\sigma_{L_{\rm \gamma}}^{2}}],
\end{equation}
where $A_{\rm L}$ is a normalized constant, $L_{\rm c}$ is the
center value and $\sigma_{L_{\rm \gamma}}$ is the width of the
distribution. The intrinsic distribution of $\theta_{\rm j}$ is
unknown. We also take it as a log-normal distribution \footnote{The
case of a broken power-law and a power-law distributions for $L_{\rm
\gamma}$ and $\theta_{\rm j}$, respectively, is discussed in section
4.},
\begin{equation}\label{DthetajG}
\psi(\theta_j)=\frac{A_\theta}{\sqrt{2\pi}\sigma_{\theta}}\exp[-\frac{(\log
\theta_j-\log \theta_{c})^{2}}{2\sigma_{\theta}^{2}}],
\end{equation}
where $A_{\rm \theta}$ is a normalized constant, $\theta_{c}$ is the
center value and $\sigma_{\theta}$ is the width of the distribution.
The observed peak flux in a given instrumental band for a GRB that
has a half-opening angle $\theta_{\rm j}$ and jet-corrected
luminosity $L_{\rm \gamma}$ at redshift $z$  is derived from
\begin{equation}\label{Pf}
P(z, L_{\rm \gamma},\theta_j)=\frac{L}{4\pi
D^{2}_{L}(z)k}=\frac{L_{\rm \gamma}}{4\pi
D^{2}_{L}(z)(1-\cos\theta_j)k},
\end{equation}
where $L=L_{\rm \gamma}/(1-\cos\theta_j)$.

The algorithm of the Swift/BAT trigger has two modes, the count rate
trigger and  image trigger (Fenimore et al. 2003; Sakamoto et al.
2008, 2011). Rate triggers are measured on different time scales (4
msec - 64 sec), with a single or several backgrounds. Image triggers
are found by summing images over various timescales and searching
for un-catalogued sources. Band (2006) compared the threshold of
Swift/BAT in count rate mode with other GRB missions. The image
trigger depends on the duration of a burst. Sakamoto et al. (2008)
parameterized the threshold of BAT as a function of burst duration.
With the trigged and off-line scanned samples of CGRO/BATSE, Qin et
al. (2010) simplified the trigger probability as a function of the
observed peak flux for BATSE. They took the same trigger probability
for BAT to make bootstrap analysis for the BAT sample. In
this analysis, we take a modified form of the trigger probability as
proposed by Qin et al. (2010).  As shown by Sakamoto et al. (2008), the trigger probability of the image trigger mode for those long GRBs with a low peak photon flux would be increased. In order to increase the trigger
probability of long and weak GRBs that have a peak flux near the BAT
threshold with the image trigger mode, we take the BAT trigger
probability as a function of $P$ as following\footnote{Note that we
focus on the redshift-known sample in this analysis. These GRBs are
usually bright. Therefore, some uncertainties of the trigger
probability of weak GRBs near the instrumental threshold would be
not significantly affected our analysis results. },
\begin{equation}\label{trig_eff}
\eta_{t}(P)=
\begin{cases}
P^{2} \  \  \  \  \  \ \  \  \  \  \  \  \ \ \ \  \ \ \ \ \ \  \  \  \  \  \  \  \ P<0.45 \  \ \  \  \  {\rm ph/ cm^{2}/ s}\\
0.67 (1.0-0.40/P)^{0.52}, \ \ P\geq0.45 \ \  \ \ \  \  {\rm ph/ cm^{2}/ s}
\end{cases}.
\end{equation}

The probability of redshift measurement for a GRB is complicated and
depends on many parameters (Fynbo et al. 2009). It is difficult to
parameterize the redshift measurement. Qin et al. (2010) got a weak
dependence of the probability on the observed peak flux , which is
quoted as following,
\begin{equation}\label{z_pro}
\eta_{z}(P)=0.26+0.032 e^{1.61\log P}.
\end{equation}
The probability of the alignment for a GRB with jet opening angle
$\theta_{\rm j}$ to an instrument with a solid angle $\Omega$ for the
aperture flux is estimated with
\begin{equation}\label{PointP}
\eta_{a}(\theta_j)=\frac{\Omega}{4\pi}(1-\cos\theta_j).
\end{equation}

Therefore, the number of GRBs with redshift measurement that trigger an instrument in an
operation period $T$ can be calculated with
\begin{equation}\label{nz}
\begin{split}
N=T\int_{0}^{\pi/2}\psi(\theta_j)d\theta_j\int_{L_{\gamma,\
\min}}^{L_{\gamma,\ \max}}
\eta_{t}(P)\eta_{z}(P)\eta_{a}(\theta_j)\Phi(L_{\rm \gamma})dL_{\rm \gamma} \\
\times \int_{0}^{z_{\max}}\frac{R_{GRB}(z)}{1+z}\frac{dV(z)}{dz}dz,
\end{split}
\end{equation}
where $z_{\max}$ is determined by the instrumental threshold of the
peak flux ($P_{\rm th}$) for a given burst with luminosity $L_{\rm
\gamma}$ according to equation (11). The $L_{\gamma,\min}$ and
$L_{\gamma,\max}$ in our analysis are taken as $10^{46}$ and
$10^{52}$ erg s$^{-1}$, respectively.
\subsection{Reference Sample for Constraining Model Parameters}
Free parameters in our models are $\rho_0$, $L_{\rm c}$ (and its
deviation $\sigma_{L_{\rm \gamma}}$), $\theta_c$ (and its deviation
$\sigma_\theta$) and $\epsilon$.  We take the current {\em
Swift}/BAT GRB sample as a reference sample to constrain these
parameters. {\em Swift}/BAT has detected 670 GRBs by June 2011.
Among them 170 GRBs have redshift measurement.  We bootstrap a mock
{\em Swift} GRB sample and compare it with the observations. The
peak fluxes, redshift and the spectral information for these GRBs
are taken from published BAT catalogs (Sakamoto et al. 2008;
Sakamoto et al. 2009; Sakamoto et al. 2011). The peak fluxes
are measured in 1-s timescale. %The bolometric peak luminosity of
%each GRB is calculated with $L=4\pi d^{2}_{L}F k$, where $F$ is the
%observed peak flux.
Note that the low-luminosity GRBs ($L<10^{49}$
erg s$^{-1}$) are not included since they may belong to a distinct
population (Soderberg et al. 2004; Liang et al. 2007; Chapman et
al. 2007).

\subsection{Bootstrapping Procedure}
(1)Generate a mock GRB characterized by $z$, $L_{\rm \gamma}$, and
$\theta_{\rm j}$. These parameters are randomly from the probability
distributions of Eqs. (6), (9), and (10) for a given set of model
parameters ($L_{\rm c}$, $\sigma_{L_{\rm \gamma}}$, $\theta_c$,
$\sigma_\theta$, $\epsilon$). Since $z<10$ for the current BAT
sample, the range of $z$ for our analysis is from 0 to 10.

(2)Calculate the isotropic gamma-ray luminosity with $L=L_{\rm
\gamma}/(1-\cos\theta_j)$ and derive $E^{'}_{p}$ in the burst frame
with an $L-E^{'}_{p}$ relation, i.e. $E^{'}_{p}=200\ {\rm keV}
(L/{10^{52}})^{1/2}/C$, where $C$ is randomly distributed in [0.1,
1] (Liang et al. 2004).

(3)Calculate the peak flux in the {\em Swift}/BAT band with Eq.
\ref{Pf} assuming that the photon indices
in the energy bands lower and higher $E_{\rm p}$ are -1 and -2.25, respectively (Preece et
al. 2000), and compare it with the threshold of {\em Swift}/BAT. A mock GRB is recognized as detectable if its peak flux is over the
threshold.

(4)Calculate the detection probability of a detectable GRB with the probabilities described by Equations
(\ref{trig_eff}), (\ref{z_pro}) and (\ref{PointP}).

$(5)$ Repeat the above steps to make a mock redshift-known BAT GRB
sample (170 GRBs).

(6)Evaluate the consistency between the mock and the observed sample
by comparing the $z$, $L$, and $\log N-\log P$ distributions. We
measure the consistency with the K-S test. We calculate the
probabilities of the K-S tests, $P_{K-S}^{P}$, $P_{K-S}^{L}$ and
$P_{K-S}^{z}$ , for the $\log N-\log P$, $L$ and $z$ distributions
and then define the global $K-S$ test probability as
$P_{K-S}^{G}=P_{K-S}^{P}\times P_{K-S}^{z}\times P_{K-S}^{L}$. A
larger value of $P_{K-S}$ indicates a better consistency. A value of
$P_{K-S}>0.1$ is generally acceptable to claim the statistical
consistency, whereas a value of $P_{K-S}<10^{-4}$ convincingly
rejects the hypothesis of the consistency.

\subsection{Results}
\begin{figure}
\includegraphics[angle=0,scale=1]{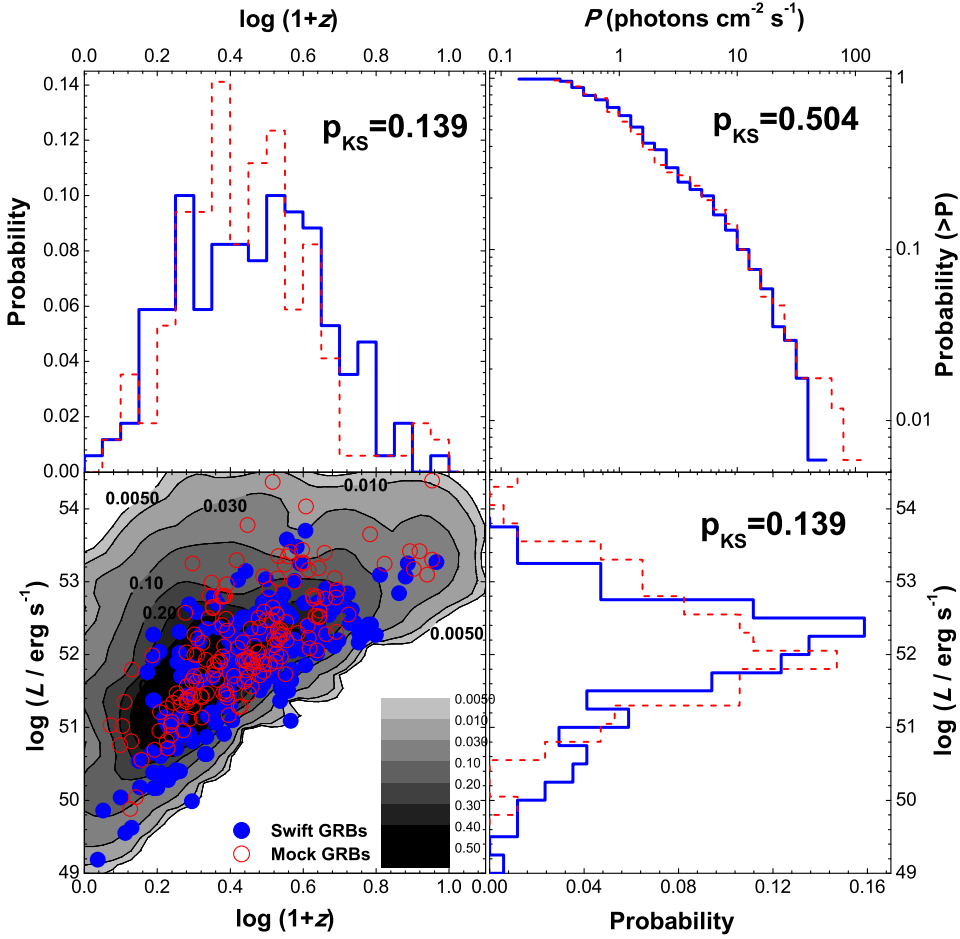}
\caption{Comparisons of two-dimensional log$L$-log$(1+z)$ distributions
and one-dimensional log $L$, log $(1+z)$ and log $P$ (accumulative
log$N$-log$P$) distributions between the 170 Swift GRBs (solid dots
and solid lines) and our results (open dots and dashed lines).
The contours in the log$L$-log $(1+z)$ plane show the normalized
probability distributions of the mock GRBs with different color
scale lines. One dimensional K-S test probabilities for the
comparisons are marked in each panel.}\label{L-z}
\end{figure}

Following the procedure described above we search for the model
parameters that can present the best consistency between the mock
and the observed BAT GRB samples.  In order to reduce the amount of
calculation, we first randomly generate a large sample of the
parameter sets($L_{\rm c}$, $\sigma_{L_{\rm \gamma}}$, $\theta_c$,
$\sigma_\theta$, $\epsilon$) in wide parameter ranges. Please note
that we do not need to consider $\rho_0$ in our bootstrap analysis
since we make the comparisons between our results and observations
with the relative probability distributions.  We find that
($\sigma_{L_{\rm \gamma}}$,$\sigma_\theta$, $\epsilon$)=(0.4, 0.6,
0.4) can well reproduce the scattering of the GRB distributions in
the $L-(1+z)$ plane. We therefore fix the three parameters and then
refine our search for the parameters $L_{\rm c}$ and $\theta_c$.
Figure (\ref{contours}a) shows the contours of the relative
$P^G_{K-S}$ in the $\log L_{\rm c}-\log \theta_c$ plane. From Figure
(\ref{contours}a) one can find that both $\theta_c$ and $L_{\rm c}$
seem to be degenerate. If $\theta_c \propto L_{\rm c}^{0.5}$, our
results roughly reproduce the observations. This is reasonable based
on the equation (\ref{Pf}). It shows that for a given instrumental
threshold $P_{\rm th}$, a lower-$L_{\rm \gamma}$ GRB may be
detectable if its $\theta_{\rm j}$ is smaller. Therefore, one cannot
constrain the intrinsic $L_{\rm c}$ and $\theta_c$ based on a
flux-limit sample. The parameter set $(\log L_{\rm c}/{\rm erg\
s}^{-1},\ \log \theta_c/{\rm rad})=(49.69, -1.27)$ gives the best
consistency to the current {\em Swift}/BAT sample. Figure \ref{L-z}
displays the comparisons between our mock sample with the BAT sample
in the $\log L-\log (1+z)$ plane along with the distributions of
$\log N-\log P$, $L$, and $z$. The $P_{K-S}$ values for the $\log
L$, $\log (1+z)$ and $\log N-\log P$ distributions are also marked
in each panel. One can observe that the observed BAT sample can be
well reproduced with our bootstrapping  method. In order to get a
robust results, we also generate a large sample of $10^4$ GRBs with
the best parameter set and make the contours of the relative
probability distribution of the GRBs in the $\log L-\log (1+z)$
plane. The result is also shown in Figure \ref{L-z}, confidently
suggesting a good consistency between our our results and
observations.

\begin{figure}
\includegraphics[angle=0,scale=1]{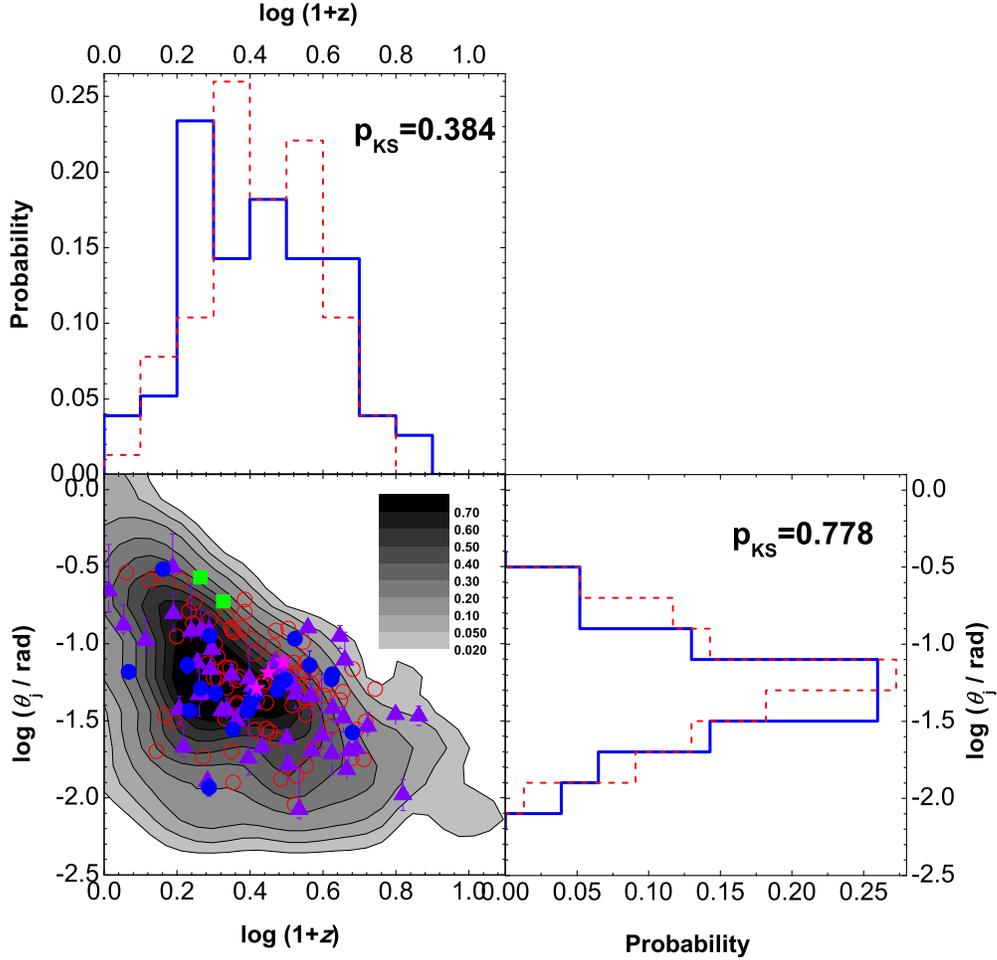}
\caption{Comparisons of two-dimensional log$\theta_{\rm
j}$-log$(1+z)$ distributions and one-dimensional log $\theta_{\rm
j}$ and log $(1+z)$ distributions between the 77 GRBs with opening
angles (solid lines) and our results (open dots and dashed lines).
The contours in the log$\theta_{\rm j}$-log $(1+z)$ plane show the
normalized probability distributions of the mock GRBs with different
color scale lines. One dimensional K-S test probabilities for the
comparisons are marked in each panel. Other symbols in the
log$\theta_{\rm j}$-log $(1+z)$ plane are the same as in Fig.
(\ref{thetaj}) }\label{thetaj-z}
\end{figure}

The $\theta_j-$known sample has 77 GRBs. We randomly pick up a
sub-sample of 77 GRBs from our mock GRB sample and show the 1- and
2-dimensional $\log \theta_j-\log (1+z)$ GRB distributions in Figure
\ref{thetaj-z}. The K-S test probabilities are also marked in each
panel in the figure. The K-S test yields a global K-S probability
larger than 0.1, indicating a good consistency between the mock and
observed BAT samples. As described in \S 3, there is no any
intrinsical relation between $\theta_{\rm j}$ and $z$ in our model.
The observed $\theta_{\rm j}-z$ dependence in the mock sample
definitely is due to instrumental selection effects. Therefore, the
apparent $\theta_{\rm j}-z$ dependence observed in the current GRB
sample would be explained with the instrumental selection effect.

As shown in Figure (\ref{contours}a), both $\theta_c$ and $L_{\rm
c}$ are degenerate. The triggered GRB number with an instrument in a
given operation period may place constraint on the $\theta_c$ and
$L_{\rm c}$, if the local GRB rate is known. However, the $\rho_0$
is quite uncertain. It is reported that $\rho_0\sim 1{\rm\
Gpc}^{-3}{\rm yr}^{-1}$ (Schmidt 2001; Lloyd-Ronning et al. 2004;
Guetta et al. 2005; Liang et al. 2007; Wanderman \& Piran 2010).
Please note that this rate is obtained without considering the jet
collimation. Assuming a typical $\theta_{\rm j}$ of 0.1 rad, the
$\rho_0$ is $\rho_0\sim 180\ {\rm Gpc}^{-3}{\rm yr}^{-1}$. The local
Ib/c SN rate, which is reported as $4.8\times 10^4 {\rm\
Gpc}^{-3}{\rm yr}^{-1}$ (Marzke et al. 1998; Cappellaro et al. 1999;
Folkes et al. 1999), presents a robust upper limit to $\rho_0$. BAT
triggered 670 GRBs in during 6.5 operation years. We therefore
calculate $\rho_0$ with equation \ref{nz} using the best parameter
sets derived above. We get $\rho_0=2.85\times10^{2} {\rm\
Gpc}^{-3}{\rm yr}^{-1}$ for the parameter set $(\log L_{\rm c}/{\rm
erg\ s}^{-1},\ \log \theta_c/{\rm rad})=(49.69, -1.27)$. The derived
local GRB rates are higher than that reported in literature, but
still under the local Ib/c SN rate. Our results suggest that $\sim
0.59\%$ of Type Ib/c SNe may be accompanied by a GRB. This may has
profound implications for understanding the relation between GRBs
and Type Ib/c SNe (e.g., Lamb et al. 2005).

\section{Conclusions and Discussion}
Assuming that the intrinsic distributions of both $L_{\rm \gamma}$
and $\theta_{\rm j}$ are Gaussian, we have shown that the observed
$\theta_{\rm j}$ dependence on redshift can be interpreted with the
instrumental selection effects.

\begin{figure}
\includegraphics[angle=0,scale=0.6]{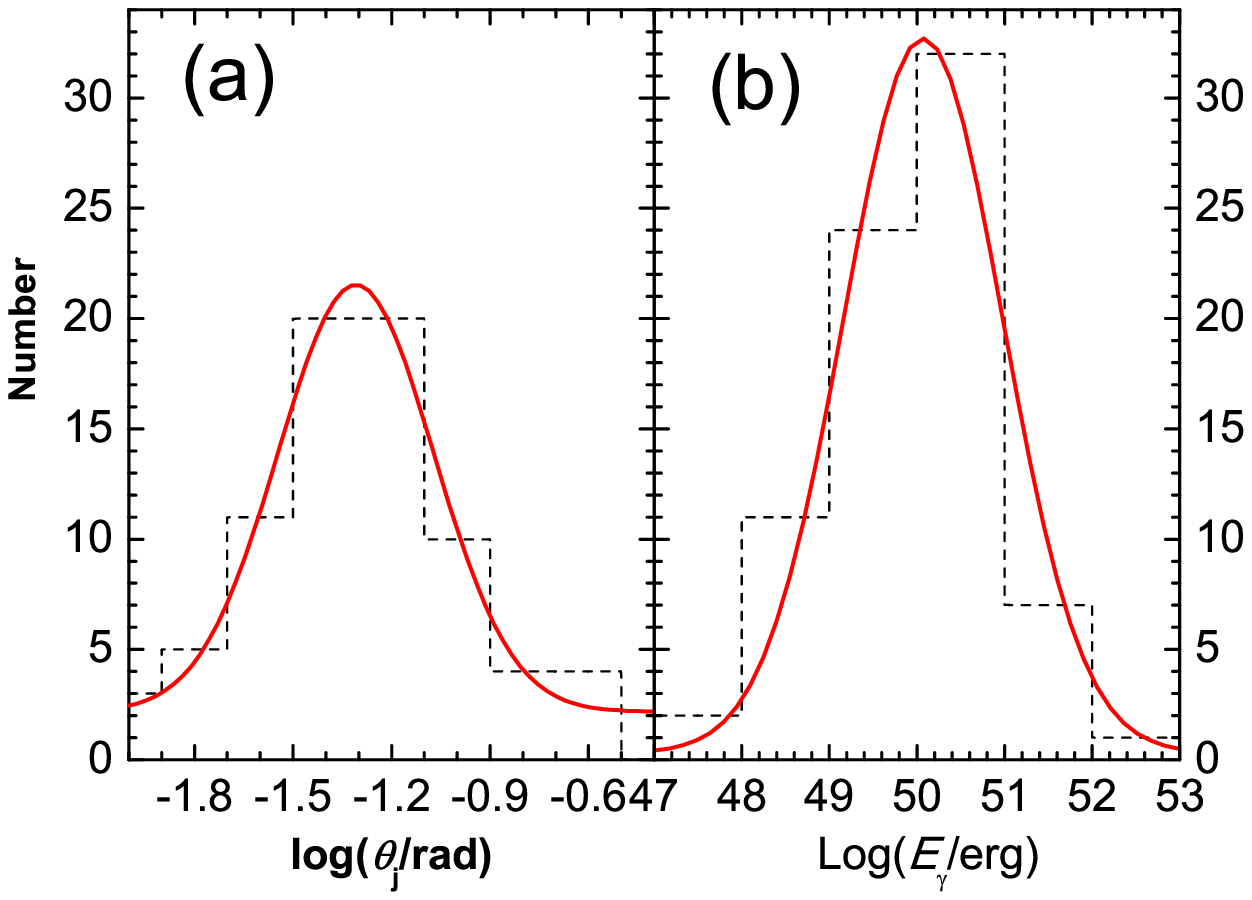}
\caption{Distributions of the jet opening angle (panel a) and the
geometrically-corrected gamma-ray energy $E_{\gamma}$ (panel b) for
the 77 GRBs in Table 1. The solid lines are the best fits to the
data with a Gaussian function.}\label{DJEr}
\end{figure}

\begin{figure}
\includegraphics[angle=0,scale=1]{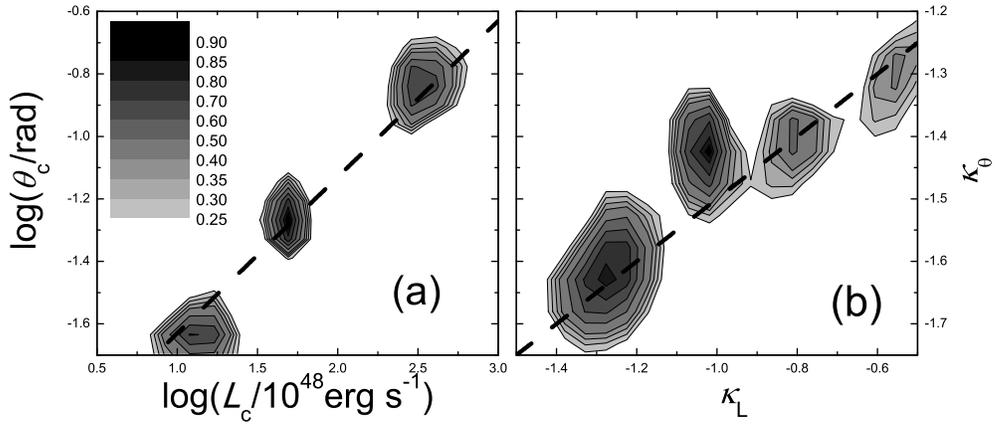}
\caption{Contours of the global K-S test probability in the $L_{\rm
c}-\theta_c$ plane (panel a) and $\kappa_{L}-\kappa_{\theta}$ plane
(panel b). The dash lines denote $\theta_c \propto L_{\rm c}^{0.5}$
and $\kappa_{\theta} \propto \kappa_{L}^{0.5}$,
respectively.}\label{contours}
\end{figure}

\begin{figure}
\includegraphics[angle=0,scale=1]{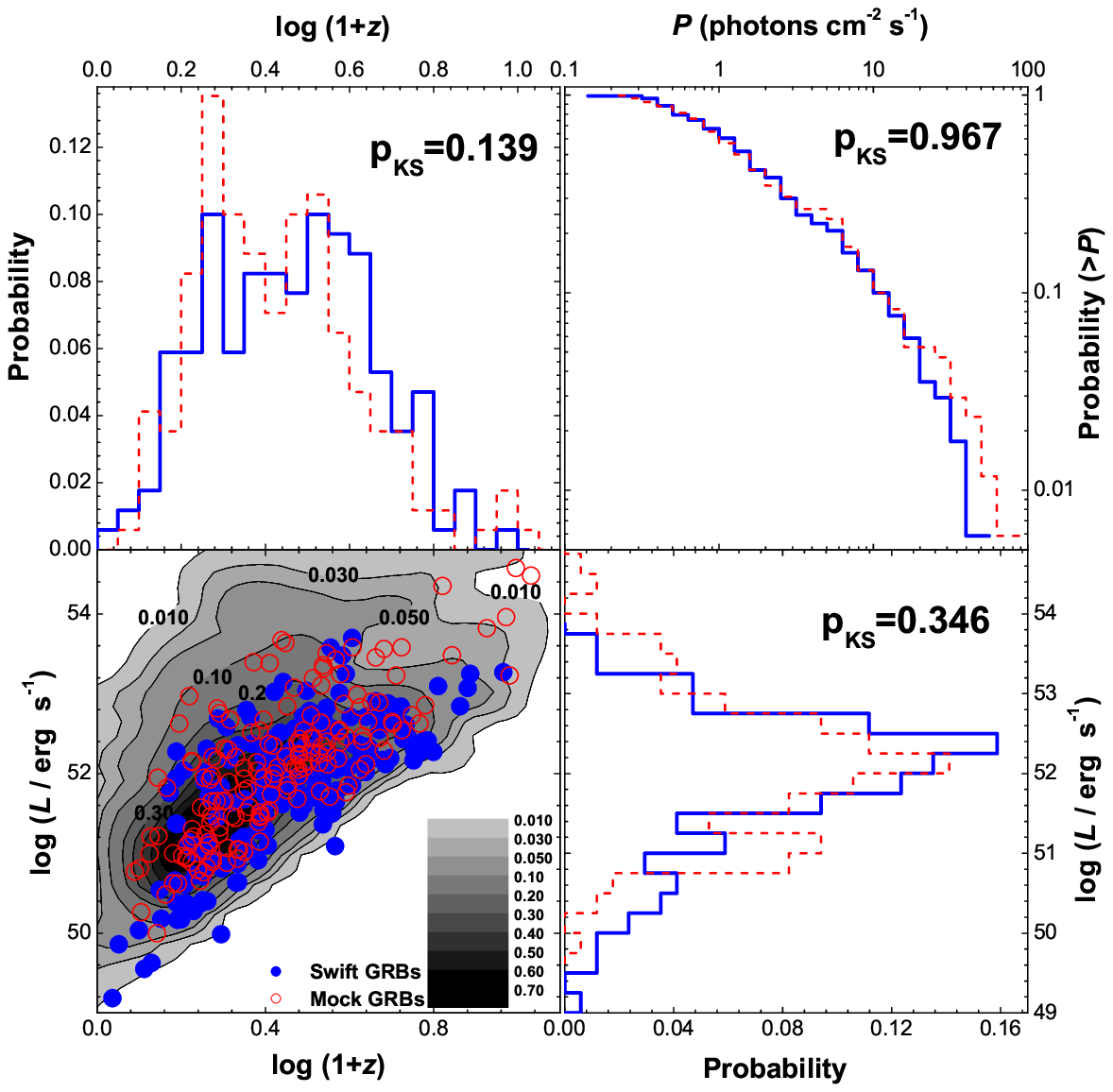}
\caption{The same as Fig \ref{L-z} but for power law distributions
of the jet-corrected luminosity and jet opening angle.}\label{L-z2}
\end{figure}

\begin{figure}
\includegraphics[angle=0,scale=0.8]{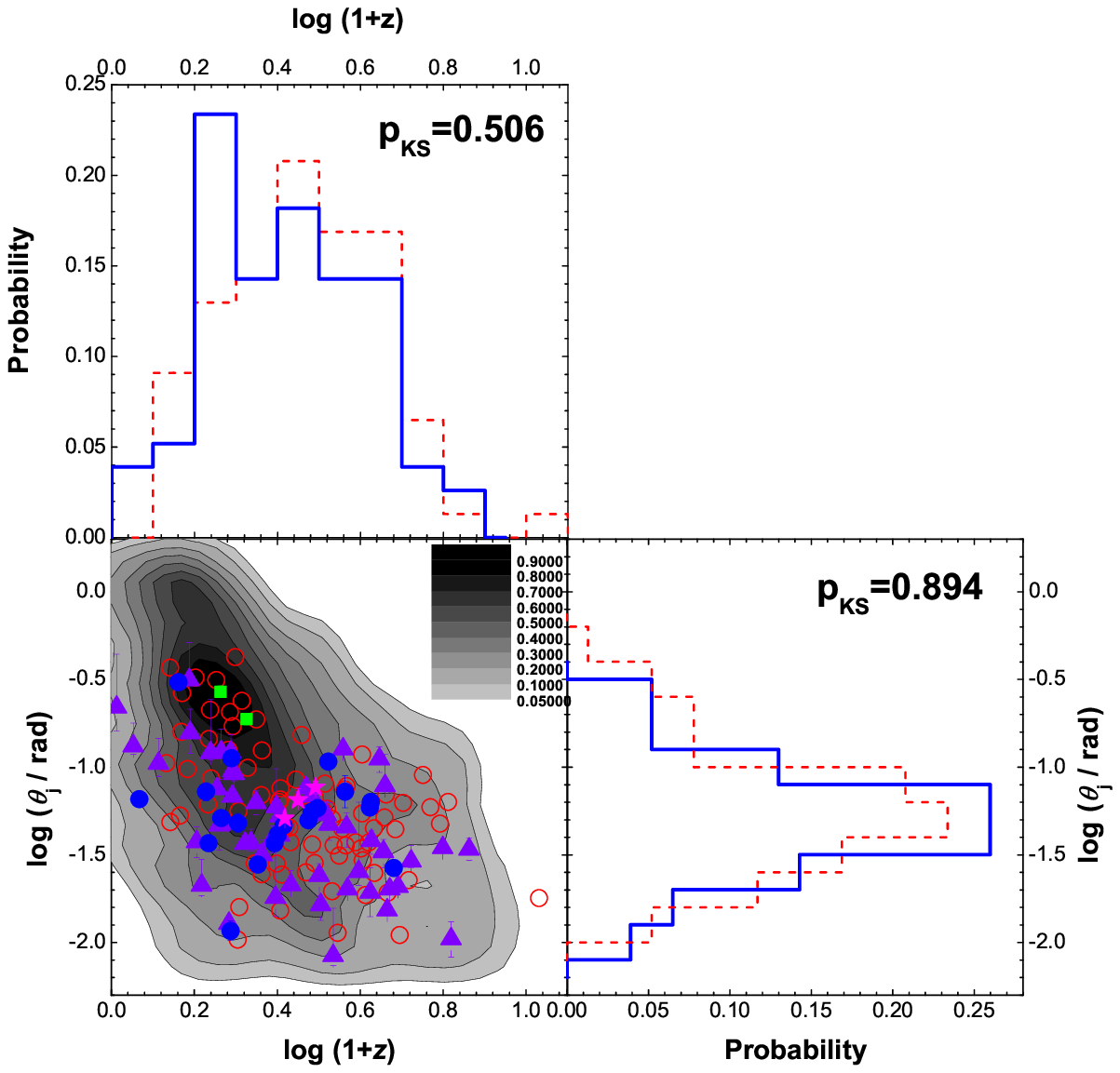}
\caption{The same as Fig \ref{thetaj-z} but for the power law
distributions of the jet-corrected luminosity and jet opening
angle.}\label{thetaj-z2}
\end{figure}

The $L_{\rm \gamma}$ (or $E_{\rm \gamma}$) and $\theta_{\rm j}$ are
two basic characteristics of the GRB jets. As shown in Figure
\ref{DJEr}, the observed distributions of $\theta_{\rm j}$ and
$E_{\rm \gamma}$ are log-normal. Frail et al. (2001) proposed the
idea of a standard energy reservoir in GRB jets based on the narrow
distribution of $E_{\rm \gamma}$ (see also Berger et al. 2003).
Although our $E_{\rm \gamma}$ distribution is still log-normal, the
width of the distribution is much larger than that reported by Frail
et al. (2001). Physically, it is difficult to access the intrinsic
distributions of both $L_{\rm \gamma}$ and $\theta_{\rm j}$ from
observations since the observed GRB sample with redshift measurement
is heavily suffered from instrumental selection effects  and
observational biases (e.g., Bloom 2003). It was suggested that  the
intrinsic distributions of $L_{\rm \gamma}$ and $\theta_{\rm j}$ are
a power-law or a broken power-law (e.g., Lin et al. 2004). As shown
by Qin et al. (2010), in case of a smooth broken power-law of
isotropic peak luminosity $L$ bootstrap analysis can well reproduce
the observed $L-z$ distributions in both the one- and
two-dimensional planes. Intuitively, being due to both instrumental
flux-limit and low triggered probability for low $L_{\rm \gamma}$,
the trigger probability for GRBs at low $L_{\rm \gamma}$ end would
be reduced significantly. Similarly, the probability of a narrower
jet pointing to the light of sight would be significantly lowered.
Hence its detection probability would rapidly drop. Therefore,
observationally, one cannot distinguish a Gaussian or a power-law
distribution for with current GRB samples. In order to examine this
expectation, we take the distributions of $L_{\rm \gamma}$ and
$\theta_{\rm j}$ as a power-law function, e.g., $\Phi(L_{\rm
\gamma})=(L_{\rm \gamma}/L^{\rm PL}_{c})^{\kappa_{L}}$;
$\psi(\theta_j)=({\theta_j}/{\theta^{\rm PL}_c})^{\kappa_{\theta}}$.
We fix $L^{\rm PL}_{c}=10^{50}{\rm erg\ s}^{-1}$ and
$\theta^{PL}_c=0.1$ rad, but let $\kappa_{L}$ and $\kappa_{\theta}$
as free parameters. Our bootstrapping method still can well
reproduce the observations. The contours of the relative
$P^{G}_{K-S}$ in the $\kappa_{L}-\kappa_{\theta}$ plane are also
plotted in Figure (\ref{contours}b) for comparison with the
$P^{G}_{K-S}$ distribution in the $\theta_c-L_{\rm c}$ plane.
Similarly, $\kappa_{L}$ and $\kappa_{\theta}$ seem to be degenerated
as that of $\theta_c$ and $L_{\rm c}$. In case of $\kappa_{\theta}
\propto \kappa_{L}^{0.5}$, our results roughly reproduce the
observations. The parameter set
$(\kappa_{L},\kappa_{\theta})=(-1.02, -1.42)$ gives the best
consistency between our results and the BAT GRB sample, which is
shown in Figs. \ref{L-z2} and \ref{thetaj-z2}. Yonetoku et al.
(2005) suggested a $\kappa$ of $ \sim -2$ from {\em CGRO}/BATSE
data, which is steeper than our result for the BAT sample. The
derived $\rho_0$ in the parameter set
$(\kappa_{L},\kappa_{\theta})=(-1.02, -1.42)$ is $1.52\times10^{2}$
${\rm Gpc}^{-3}{\rm yr}^{-1}$.

Some apparent correlations between some GRB properties and redshift
were discussed by some authors with small samples, such as the
redshift dependence of the spectral lag of GRBs in prompt phases (Yi
et al. 2008) and the duration of the shallow decay phase of GRB
X-ray afterglows (Stratta et al. 2009). Physically, these
correlations may be results from the cosmic evolution of the
explosion energy and/or the jet geometry (e.g., Xu et al. 2005), if
they are intrinsic. However, as shown in our analysis, there are no
convincing cosmic evolution features. Therefore, these apparent
correlations may be, at least partially, due to sample selection
effects or observational biases.

Tight correlations between $E_{\rm \gamma}$ and $E_p^{'}$ (Ghirlanda
relation; Ghirlanda et al. 2004) or among $E_{\rm iso}$, $t_{\rm j}$
and $E_p^{'}$ (Liang-Zhang relation, Liang \& Zhang 2005) were
discovered with small GRB sample. As shown in this analysis, the
$E_{\gamma}$ distribution would  suffer great selection effects. It
is known that the observed distribution of $E_{\rm p}$ is also due
to the selection effect. The observed narrow distribution of $E_{\rm
p}$ by CGRO/BATSE is significantly broadened by the observations
with HETE-2 and BAT. Our analysis cannot figure out if the
selections effects on both $E_{\rm \gamma}$ and $E_{\rm p}$
distributions can be canceled out and evaluates whether these
relations are intrinsic.

\acknowledgments We acknowledge the use of the public data from the
Swift data archive. We thank helpful discussion with Bing Zhang, X.
F. Wu, and Shuang-Nan Zhang. This work is supported by the National
Natural Science Foundation of China under grants No. 11063001,
10873002 and 11025313, the National Basic Research Program (``973"
Program) of China (Grant 2009CB824800), Special Foundation for Distinguished Expert Program of Guangxi
, the Guangxi SHI-BAI-QIAN project (Grant 2007201), the Guangxi Natural Science Foundation
(2010GXNSFA013112 and 2010GXNSFC013011), the special funding for national outstanding young scientist (Contract No. 2011-135), and the 3th Innovation Projet of Guangxi University.

%\clearpage

\clearpage
\begin{deluxetable}{llllllllll}
%\begin{longtable}{llllllllll}
\tablewidth{440pt}
 \tabletypesize{\footnotesize}
\tabletypesize{\tiny}
\tablecaption{Jet Break Times and Jet Opening Angles} \tablenum{1}
\tablehead{\colhead{GRB}&\colhead{$z$}&\colhead{$S_{\gamma}$}&\colhead{Energy$^{a}$}&\colhead{$E_{\rm iso}(\gamma)^{b}$}&\colhead{$t_{\rm j}$}&\colhead{Band$^{c}$}&\colhead{$\theta_{\rm j}^{d}$}&\colhead{$E_{\gamma}^{e}$}&\colhead{Refs.}
\\  \colhead{}& \colhead{}&\colhead{($10^{-6}$ erg cm$^{-2}$)}&\colhead{(keV)}&\colhead{($10^{51} $erg)}&\colhead{(days)}&\colhead{}&\colhead{(rad)}&\colhead{($10^{51} $erg)}&\colhead{} }
\startdata
970508  &   0.835   &   3.17    &   20-2000   & 8.25      &   25$\pm$5        &   R   &   0.269$\pm$0.020    & 0.296 & 1,2,3   \\
970828  &   0.958   &   96.00   &   20-2000   & 272       &   2.2$\pm$0.4     &   X   &   0.068$\pm$0.005   & 0.631 & 4,2,4   \\
980703  &   0.966   &   22.60   &   20-2000   & 67.2      &   3.4$\pm$0.5     &   X &   0.095$\pm$0.005   & 0.306 & 5,2,6   \\
990123  &   1.60    &   268.00  &   20-2000   & 1945      &   2.04$\pm$0.46   &   O   &   0.047$\pm$0.004   & 2.11  & 7,2,7   \\
990510  &   1.619   &   22.60   &   20-2000   & 182       &   1.20$\pm$0.08   &   O/R &   0.051$\pm$0.001   & 0.239 & 8,2,9   \\
990705  &   0.84    &   93.00$\pm$2.00  &   40-700   &  298   &   1.0$\pm$0.2 &   O   & 0.051$\pm$0.004 & 0.393 & 10,11,12    \\
991216  &   1.02    &   194.00  &   20-2000   & 655  &   1.2$\pm$0.4     &   O   &   0.048$\pm$0.006   & 0.756 & 13,2,14 \\
000301C &   2.034   &   4.10    &   25-1000   & 57.7  &   7.3$\pm$0.5     &   O &   0.11$\pm$0.003    & 0.349 & 15,16,16    \\
000418  &   1.119   &   20.00   &   15-1000   & 95.3  &   25$\pm$5        &   R   &   0.188$\pm$0.014   & 1.67  & 17,18,18    \\
000926  &   2.037   &   6.20    &   25-100    & 200   &   1.8$\pm$0.1     &   O   &   0.056$\pm$0.001   & 0.310 & 19,20,21    \\
010222  &   1.477   &   120.00$\pm$3.00    &    2-700  &    1383            &   $0.93^{+0.15}_{-0.06}$  &   O   &  $0.037^{+0.002}_{-0.0009}$  & 0.940 &    22,23,24    \\
010921  &   0.451   &   15.40$\pm$0.20     &    8-400  &    13.8            &   33.0$\pm$6.5            &   O   &  0.306$\pm$0.023             & 0.638 &    25,26,27    \\
011211  &   2.14    &   5.00    &   40-700   &  117   &   1.77$\pm$0.28   &   O   &   0.058$\pm$0.004   & 0.199 & 28,29,30    \\
020124  &   3.2     &   6.10    &   30-400   &  227   &   3.0$\pm$0.4     &   O   &   0.059$\pm$0.003   & 0.392 & 31,31,32    \\                                                                                   020405  &   0.69    &   38.00$\pm$4.00     &    50-700   &   114           &   1.67$\pm$0.52     &    O   &  0.072$\pm$0.009  & 0.299 &   25,33,30    \\
020813  &   1.254   &   38.00   &   25-100   &  1701   &   0.43$\pm$0.06   &   O   &   0.028$\pm$0.002     &   0.661 & 34,35,34    \\
021004  &   2.332   &   3.20    &   7-400    &  58.9   &   7.6$\pm$0.3     &   O   &   0.108$\pm$0.002     &   0.341 & 36,37,38    \\
030226  &   1.986   &   5.70    &   30-400   &  98.9   &   1.04$\pm$0.12   &   O   &   0.050$\pm$0.002     &   0.123 & 39,39,39    \\
030328  &   1.52    &   29.00   &   30-400   &  350   &   0.8$\pm$0.1     &   O   &   0.041$\pm$0.002     &   0.295 & 31,31,40    \\
030329  &   0.168   &   120.00  &   30-400   &  16.8   &   0.47$\pm$0.005  &   O   &   0.066$\pm$0.0009    &   0.036 & 41,41,41    \\
030429  &   2.66    &   0.38    &   30-400   &  13.2   &   1.77$\pm$1.0    &   O   &   0.073$\pm$0.015     &   0.035 & 31,31,42    \\
041006  &   0.716   &   7.00    &   30-400   &  21.0   &   0.16$\pm$0.04   &   O   &   0.037$\pm$0.003     &   0.014 & 29,29,43    \\
050315  &   1.95    &   3.22$\pm$0.146  &   15-150   &  67.4   &   $2.78^{+0.80}_{-0.88}$  &   X   &   $0.076^{+0.008}_{-0.009}$   &   0.195 & 44,44,45    \\
050318  &   1.44    &   1.08$\pm$0.077  &   15-150   &  17.6   &   $0.24^{+0.12}_{-0.11}$  &   X   &   0.038$\pm$0.007     &   0.013 & 44,44,45    \\
050319  &   3.24    &   $1.9^{+0.2}_{-0.3}$ &   15-350 &    69.6 &   $0.64^{+0.22}_{-0.30}$  &   X   &   $0.038^{+0.005}_{-0.007}$   &   0.050 & 46,46,45    \\
050408  &   1.236   &   1.4     &   2-30     &  89.9   &   1.39$\pm$0.58       &   X   &   0.063$\pm$0.01      &   0.177 & 47,48,49    \\
050416A &   0.65    &   0.37$\pm$0.04   &   15-150    & 0.991  &   $0.01^{+0.01}_{-0.005}$     &   X   &   $0.021^{+0.007}_{-0.0.003}$     &   0.0002 &    44,44,45    \\
050505  &   4.28    &   2.49$\pm$0.18   &   15-150    & 174  &   $0.53^{+0.29}_{-0.13}$      &   X   &   $0.029^{+0.006}_{-0.003}$   &   0.074 & 44,44,45    \\
050525A &   0.606   &   15.30$\pm$0.22  &   15-150    & 22.6  &   0.16$\pm$0.09   &   X   &   0.037$\pm$0.008 &   0.016 & 44,44,50    \\
050802  &   1.71    &   2.00$\pm$0.16   &   15-150    & 37.0  &   $0.07^{+0.04}_{-0.01}$  &   X   &   $0.021^{+0.004}_{-0.002}$   &   0.008 & 44,44,45    \\
050814  &   5.3     &   2.01$\pm$0.22   &   15-150    & 184  &   $1.03^{+0.26}_{-0.18}$  &   X   &   $0.035^{+0.003}_{-0.002}$   &   0.111 & 44,44,45    \\
050820A &   2.62    &   3.44$\pm$0.24   &   15-150    & 163  &   18$\pm$2        &   X   &   0.127$\pm$0.005 &   1.31 &  44,44,51    \\
050826  &   0.3     &   0.41$\pm$0.07   &   15-150    & 0.242  &   $0.45^{+0.39}_{-0.20}$  &   X   &   $0.105^{+0.035}_{-0.018}$   &   0.001 & 44,44,45    \\
050904  &   6.29    &   5.40$\pm$0.20   &   15-150    & 2267  &   2.6$\pm$1       &   O   &   0.034$\pm$0.005         &   1.31 &  52,52,52    \\
050922C &   2.2     &   1.62$\pm$0.05   &   15-150    & 56.5  &   $0.05^{+0.01}_{-0.03}$  &   X   &   $0.016^{+0.002}_{-0.003}$   &   0.008 & 44,44,45    \\
051016B &   0.94    &   0.17$\pm$0.02   &   15-150    & 0.969  &   $1.56^{+0.96}_{-1.00}$  &   X   &   $0.122^{+0.028}_{-0.029}$   &   0.007 & 44,44,45    \\
051022  &   0.8     &   160     &   2-400    &  362   &   2.9$\pm$0.2     &   X   &   0.075$\pm$0.002         &   1.02 &  53,53,53    \\
051109A &   2.35    &   2.20$\pm$0.27   &   15-150    & 75.9  &   $0.92^{+0.71}_{-0.51}$  &   X   &   $0.047^{+0.014}_{-0.001}$   &   0.084 & 44,44,45    \\
051111  &   1.55    &   4.08$\pm$0.13   &   15-150    & 81.4  &   $0.49^{+0.16}_{-0.10}$  &   X   &   $0.041^{+0.005}_{-0.003}$   &   0.068 & 44,44,45    \\
051221A &   0.55    &   2.40$\pm$0.40   &   100-2000  & 4.46  &   $4.07^{+3.40}_{-2.88}$  &   X   &   $0.157^{+0.049}_{-0.042}$   &   0.055 & 44,54,45    \\
060115  &   3.53    &   1.71$\pm$0.15   &   15-150    & 92.3  &   $0.51^{+0.22}_{-0.28}$  &   X   &   $0.033^{+0.005}_{-0.007}$   &   0.050 & 44,44,45    \\
060124  &   2.3     &   0.48$\pm$0.03   &   15-350    & 11.9  &   $0.68^{+0.32}_{-0.14}$  &   X   &   $0.053^{+0.009}_{-0.004}$   &   0.017 & 44,46,45    \\
060206  &   4.05    &   0.83$\pm$0.04   &   15-150    & 57.9  &   0.57$\pm$0.06   &   O &   0.035$\pm$0.001 &   0.035 & 44,44,55    \\
060210  &   3.91    &   7.66$\pm$0.41   &   15-150    & 569  &   $0.30^{+0.11}_{-0.08}$  &   X   &   $0.021^{+0.003}_{-0.002}$   &   0.123 & 44,44,45    \\
060218  &   0.03    &   1.57$\pm$0.15   &   15-150    & 0.008  &   $0.82^{+1.53}_{-0.69}$  &   X   &   $0.220^{+0.153}_{-0.069}$   &   0.0002 &    44,44,45    \\
060418  &   1.49    &   8.33$\pm$0.25   &   15-150    & 144  &   $0.07^{+0.20}_{-0.04}$  &   X   &   $0.018^{+0.020}_{-0.005}$   &   0.024 & 44,44,45    \\
060526  &   3.22    &   1.26$\pm$0.15   &   15-150    & 63.7  &   2.41$\pm$0.06   &   O   &   0.063$\pm$0.001 &   0.128 & 44,44,56    \\
060605  &   3.8     &   0.70$\pm$0.09   &   15-150    & 43.6  &   0.24$\pm$0.02   &   O   &   0.027$\pm$0.001 &   0.016 & 44,44,57    \\
060614  &   0.13    &   28.20$\pm$0.40  &   15-350    & 2.04  &   $1.45^{+1.16}_{-0.41}$  &   X   &   $0.132^{+0.040}_{-0.014}$   &   0.018 & 44,46,45    \\
060707  &   3.43    &   1.60$\pm$0.15   &   15-150    & 79.9  &   $12.26^{+5.25}_{-5.72}$ &   X   &   $0.111^{+0.018}_{-0.020}$   &   0.495 & 44,44,45    \\
060714  &   2.71    &   2.83$\pm$0.17   &   15-150    & 109  &   0.12$\pm$0.01   &   X   &   0.020$\pm$0.001 &   0.022 & 44,44,58    \\
060729  &   0.54    &   2.61$\pm$0.21   &   15-150    & 4.70  &   $26.23^{+34.61}_{-6.12}$    &   X   &   $0.314^{+0.155}_{-0.028}$   &   0.229 & 44,44,45    \\
060814  &   0.84    &   14.60$\pm$0.24  &   15-150    & 88.7  &   0.55$\pm$0.14   &   X   &   0.048$\pm$0.005 &   0.101 & 44,44,45    \\
060906  &   3.69    &   2.21$\pm$0.14   &   15-150    & 138  &   $0.16^{+0.04}_{-0.03}$  &   X   &   $0.020^{+0.002}_{-0.001}$   &   0.028 & 44,44,45    \\
060908  &   2.43    &   2.80$\pm$0.11   &   15-150    & 101  &   $0.01^{+0.02}_{-0.003}$ &   X   &   $0.008^{+0.005}_{-0.001}$   &   0.004 & 44,44,45    \\
060926  &   3.21    &   0.22$\pm$0.03   &   15-150    & 11.5  &   0.06$\pm$0.05   &   X   &   0.019$\pm$0.006 &   0.002 & 44,44,45    \\
060927  &   5.6     &   1.13$\pm$0.07   &   15-150    & 119  &   0.04$\pm$0.02   &   X   &   0.011$\pm$0.002 &   0.007 & 44,44,45    \\
061121  &   1.31    &   13.70$\pm$0.20  &   15-150    & 280  &   $0.34^{+0.18}_{-0.08}$      &   X   &   $0.032^{+0.006}_{-0.003}$       &   0.140 & 44,44,45    \\
070125  &   1.547   &   22.50$\pm$3.50  &   20-10000  & 108  &   $1.05^{+0.35}_{-0.30}$  &   X   &   $0.053^{+0.007}_{-0.006}$   &   0.149 & 59,59,45    \\
070208  &   1.17    &   0.45$\pm$0.10   &   15-150    & 3.66  &   $0.11^{+0.05}_{-0.03}$  &   X   &   $0.037^{+0.007}_{-0.004}$   &   0.003 & 44,44,45    \\
070306  &   1.5     &   $9.00^{+0.50}_{-0.80}$  &   15-350    & 88.8  &   $1.33^{+1.79}_{-0.86}$  &   X   &   $0.059^{+0.030}_{-0.014}$   &   0.156 & 44,46,45    \\
070318  &   0.84    &   2.48$\pm$0.11   &   15-150    & 13.2  &   $3.57^{+2.88}_{-0.63}$  &   X   &   $0.122^{+0.037}_{-0.008}$   &   0.098 & 44,44,45    \\
070411  &   2.95    &   2.70$\pm$0.16   &   15-150    & 115  &   $0.24^{+0.58}_{-0.11}$  &   X   &   $0.025^{+0.023}_{-0.005}$   &   0.037 & 44,44,45    \\
070508  &   0.82    &   19.60$\pm$0.27  &   15-150    & 124  &   $0.58^{+0.93}_{-0.50}$  &   X   &   $0.047^{+0.028}_{-0.015}$   &   0.136 & 44,44,45    \\
070611  &   2.04    &   0.39$\pm$0.06   &   15-150    & 9.25  &   $1.13^{+0.39}_{-0.53}$  &   X   &   $0.069^{+0.009}_{-0.012}$   &   0.022 & 44,44,45    \\
070714B &   0.92    &   0.72$\pm$0.09   &   15-150    & 19.2  &   0.01$\pm$0.002  &   X   &   0.013$\pm$0.001 &   0.002 & 44,44,45    \\
070721B &   3.63    &   7.40$\pm$0.60   &   15-350    & 365  &   $0.11^{+0.01}_{-0.02}$  &   X   &   0.015$\pm$0.001     &   0.043 & 60,60,45    \\
070810A &   2.17    &   0.69$\pm$0.06   &   15-150    & 17.3  &   $0.09^{+0.04}_{-0.06}$  &   X   &   $0.024^{+0.004}_{-0.006}$   &   0.005 & 44,44,45    \\
071003  &   1.1     &   8.30$\pm$0.30   &   15-150    & 206  &   $0.41^{+0.07}_{-0.09}$  &   X   &   $0.037^{+0.002}_{-0.003}$   &   0.139 & 44,44,45    \\
071010A &   0.98    &   0.20$\pm$0.04   &   15-150    & 1.22  &   $0.81^{+0.22}_{-0.20}$  &   X   &   $0.092^{+0.010}_{-0.009}$   &   0.005 & 44,44,45    \\
071010B &   0.947   &   4.40$\pm$0.10   &   15-150    & 19.1  &   3.44$\pm$0.39   &   O   &   0.113$\pm$0.005 &   0.121 & 44,44,61    \\
071031  &   2.69    &   0.90$\pm$0.13   &   15-150    & 34.3  &   $0.71^{+0.35}_{-0.50}$  &   X   &   $0.046^{+0.008}_{-0.012}$   &   0.036 & 44,44,45    \\
080319B &   0.937   &   81.00$\pm$1.00  &   15-150    & 1173  &   0.03$\pm$0.01   &   O   &   0.012$\pm$0.002 &   0.080 & 44,44,62    \\
090323  &   3.568   &   $202.00^{+28.00}_{-25.00}$  &   20-10000   & 3722  &  $17.80^{+19.60}_{-1.70}$  & X   &   $0.078^{+0.032}_{-0.003}$  &  11.4 &    63,63,64    \\
090328  &   0.736   &   95.00$\pm$10.00 &   8-40000   & 100  &   $6.40^{+12.00}_{-1.50}$ &   X   &   $0.121^{+0.085}_{-0.011}$   &   0.729 & 65,65,64    \\
090902B &   1.822   &   374.00$\pm$3.00 &   50-10000  & 2972  &   $6.20^{+2.40}_{-0.80}$  &   O/X &   $0.065^{+0.009}_{-0.003}$   &   6.27 &  66,67,64    \\
090926A &   2.106   &   145.00$\pm$4.00 &   8-1000    & 1808  &   9.00$\pm$2.00   &   O/X &   0.077$\pm$0.006 &   5.32 &  68,69,64    \\
\enddata
%\tablenotetext{}{Redshift (\emph{z}), fluence
%($S_{\gamma}$), and jet break times ($t_{j}$) were taken from the listed references. References are
%given in the order of the redshift, fluence, and jet-break time.}
\tablenotetext{a}{The energy range ($keV$) over which the fluence
was reported.} \tablenotetext{b}{The isotropic gamma-ray energies
($E_{\rm iso}(\gamma)$)} \tablenotetext{c}{The jet-break is
identified in radio (R), optical (O), or X-ray (X), respectively.}
\tablenotetext{d}{The jet opening angle ($\theta_{\rm j}$)}
\tablenotetext{e}{The geometrically-corrected gamma-ray energy
($E_{\gamma}$)} \tablerefs{(1) Bloom et al. 1998; (2) Piran,
Jimenez, \& Band 2000; (3) Frail et al. 2000; (4) Djorgovski et al.
2001; (5) Djorgovski et al. 1998; (6) Frail et al. 2003; (7)
Kulkarni et al. 1999; (8) Vreeswijk et al. 2001; (9) Harrison et al.
1999; (10) Le Floc'h et al. 2002; (11) Amati et al. 2000; (12)
Masetti et al. 2000; (13) Vreeswijk et al. 1999; (14) Halpern et al.
2000; (15) Castro et al. 2000a; (16) Berger et al. 2000; (17) Bloom
et al. 2003; (18) Berger et al. 2001; (19) Castro et al. 2000b; (20)
Price et al. 2001; (21) Harrison et al. 2001; (22) Mirabal et al.
2002; (23) in't Zand et al. 2001; (24) Galama et al. 2003; (25)
Price et al. 2002; (26) Ricker et al. 2002; (27) Price et al. 2003a;
(28) Holland et al. 2002; (29) Frontera et al. 2002; (30) Bloom,
Frail, \& Kulkarni 2003; (31) $http://space.mit.edu/HETE/Bursts/$;
(32) Berger, Kulkarni, \& Frail 2003; (33) Price et al. 2003b; (34)
Barth et al. 2003; (35) Hurley et al. 2002; (36) Matheson et al.
2003; (37) Lamb et al. 2002; (38) Pandey et al. 2003; (39) Klose et
al. 2004; (40) Andersen et al. 2003; (41) Willingale et al. 2004;
(42) Jakobsson et al. 2004; (43) Stanek et al. 2005; (44)
$http://heasarc.gsfc.nasa.gov/docs/swift/archive/grb\b{ }
table/index.php$; (45) Racusin et al. 2009; (46) Butler et al. 2007;
(47) Berger, Gladders, \& Oemler 2005; (48) Sakamoto et al. 2005;
(49) Covino et al. 2005; (50) Liang et al. 2008; (51) Cenko et al.
2006; (52) Tagliaferri et al. 2005; (53) Racusin et al. 2005; (54)
Endo et al. 2005; (55) Curran et al. 2007; (56) Th$\ddot{o}$ene et
al. 2010; (57) Ferrero et al. 2009; (58) Krimm et al. 2007; (59)
Nava et al. 2008; (60) Butler, Bloom, \& Poznanski 2010; (61) Kann
et al. 2007; (62) Racusin \& Burrows 2008; (63) Golenetskii et al.
2009; (64) Cenko et al. 2011; (65) Rau, Connaughton, \& Briggs 2009;
(66) Pandey et al. 2010; (67) Bissaldi \& Connaughton 2009;
(68)Malesani et al. 2009; (69) Bissaldi 2009.}

\end{deluxetable}
%\end{longtable}
\end{document}